\documentstyle[12pt,aasms4]{article}

\begin{document}

\title{A New Code for Nonradial Stellar Pulsations and its Application
to Low - Mass, Helium White Dwarfs}

\author{A.    H.    C\'orsico\altaffilmark{1}    and    O.     G.
Benvenuto\altaffilmark{2}}

\affil{Facultad   de  Ciencias  Astron\'omicas   y  Geof\'{\i}sicas,\\
Universidad Nacional de La Plata,\\  Paseo del Bosque S/N,\\ (1900) La
Plata, Argentina}

\vskip 5cm

\altaffiltext{1}{Fellow  of the  Consejo  Nacional de  Investigaciones
Cient\'{\i}ficas   y    T\'ecnicas   (CONICET),   Argentina.    Email:
acorsico@fcaglp.fcaglp.unlp.edu.ar}

\altaffiltext{2}{Member    of    the    Carrera    del    Investigador
Cient\'{\i}fico, Comisi\'on de  Investigaciones Cient\'{\i}ficas de la
Provincia    de    Buenos    Aires    (CIC),    Argentina.\\    Email:
obenvenuto@fcaglp.fcaglp.unlp.edu.ar}

\begin{abstract} We present a finite difference code intended for
computing linear, adiabatic,  nonradial pulsations of spherical stars.
This code  is based on a  slight modification of the  general Newton -
Raphson technique in order to  handle the relaxation of the eigenvalue
(square of  the eigenfrequency) of  the modes and  their corresponding
eigenfunctions.   This code  has been  tested computing  the pulsation
spectra of  polytropic spheres finding a good  agreement with previous
work.

Then, we have  coupled this code to our  evolutionary code and applied
it to the computation of the  pulsation spectrum of a low mass, pure -
helium white  dwarf of 0.3 $M_{\odot}$  for a wide  range of effective
temperatures.    In  making   this  calculation   we  have   taken  an
evolutionary time step short enough such that eigenmodes corresponding
to a  given model are  used as initial  approximation to those  of the
next  one. Specifically,  we  have computed  periods, period  spacing,
eigenfunctions,  weight functions,  kinetic  energies and  variational
periods for  a wide range of modes.   To our notice this  is the first
effort in  studying the pulsation  properties of helium  white dwarfs.
The solution  we have found  working with these realistic  white dwarf
models  are in  good accord  with  the predictions  of the  asymptotic
theory of  Tassoul (1980) for  high order modes.  This  indicates that
the code presented here is able to work adequately also with realistic
stellar models. \end{abstract}


\section{INTRODUCTION} \label{sec_introd}

At present, nonradial stellar pulsations constitute a valuable tool in
extracting information about the  properties and internal structure of
objects  that undergo  such  kind of  oscillations.  As  observational
techniques  are  refined, a  growing  variety  of  stellar objects  is
discovered that undergo  nonradial oscillations.  The most extensively
studied case  is our Sun, which has  been shown to vibrate  in a large
amount  of modes,  investigation  of  which has  been  so powerful  in
revealing  clues of its  internal structure  that it  is known  as the
discipline of  {\it Helioseismology}.   Also, it has  been established
that different kinds of stellar objects which are located in a variety
of  places in the  HR diagram,  covering several  evolutionary stages,
undergo nonradial  pulsations, e.g.:  Ap, WR, $\delta$  Scuti, $\beta$
Cephei, and variable  white dwarfs (hereafter WDs): DAV,  DBV, DOV and
PNNV.  The study  of  oscillations  in such  variable  stars, and  the
subsequent comparison and  fitting with theoretical pulsation patterns
is now known as {\it Asteroseismology}.  For a general introduction to
this  topic see,  e.g.,  Cox (1980);  Unno,  et al.  (1989); Brown  \&
Gilliland (1994); and Gaustchy \& Saio (1995; 1996).

From the  observational point of  view (with the obvious  exception of
our  Sun),  WDs  represent  the  best established  kind  of  nonradial
pulsators. WDs are the final fate for low and intermediate mass stars.
During  its cooling,  they go  across very  narrow  instability strips
inside which suffer nonradial pulsations in the branch of g- (gravity)
modes with periods  between 100 and 2000 sec and  amplitudes up to 0.3
magnitudes.

As WDs cools down, the  outermost layers main constituent (hydrogen in
DAVs, helium in DBVs, and probably carbon or oxygen in DOVs and PNNVs;
see e.g. Brown \& Gilliland 1994 and Gaustchy \& Saio 1995; 1996) gets
partial ionization conditions which forces the development of an outer
convection zone  (OCZ). In the  bottom of such zone  the instabilities
are originated ($\kappa$ mechanism of overstability).  This gives rise
to the blue  edge of the instability strip  (see, e.g., Dziembowski \&
Koester 1981 and  Winget, et al.  1982; see  also Tassoul, Fontaine \&
Winget 1990,  hereafter TFW)\footnote{Recently Goldreich  \& Wu (1999)
have proposed the ``convective  driving mechanism'' as the responsible
for overstability of g-modes in  DA WDs.}. The physical reason for the
appearance of a red edge is still under debate.

At present,  there is a general  consensus that variable  WDs are very
interesting  objects  for  asteroseismological  studies.   Their  very
simple internal structure allows us to predict theoretically the modes
with a very high degree  of detail and sophistication. Also, they have
a very  rich spectrum of periods  which may give  us information about
stellar masses (by  means of measuring the mean  period spacing), core
composition (by measuring the rate of change of the period $\dot{P}$),
mass of the surface helium  and hydrogen layers (if present) (by means
of  the departures  from uniform  period spacing  and  mode trapping),
velocity of rotation and strength  of the magnetic field (studying the
fine structure of  a multiplet), etc. (see, e.g.,  Bradley 1996).  So,
it is not surprising  that in the last years DAVs and  DBVs as well as
DOVs have  been the  preferred target for  the network  called ``Whole
Earth Telescope'' (WET). WET  observations have been of an unprecedent
quality which in  some cases (see, e.g., Winget  et al.  1991; Kawaler
1993;  Bradley \&  Winget 1994)  allowed to  apply the  powerful tools
above mentioned.

In view  of the growing  importance of this  line of research,  in our
Observatory  we have begun  the study  of nonradial  oscillations.  To
this  end, we  have developed  a  code to  compute eigenfunctions  and
eigenfrequencies  for  spherical  objects  in  the  linear,  adiabatic
approximation. The  main purpose of  the present paper is  to describe
the numerical techniques employed in  our program to search for and to
compute the modes and its application to the case of a 0.3 $M_{\odot}$
pure helium  WD. This code is  based on a modification  of the general
Newton  -  Raphson  technique  presented  in  Kippenhahn,  Weigert  \&
Hofmeister  (1967) to  solve  the set  of  difference equations  which
represent the  differential equations of  linear, adiabatic, nonradial
pulsations of spherical stars.

In order to test the code we have applied it to the case of polytropic
spheres  whose  eigenfrequencies have  been  computed  with very  high
accuracy  (Christensen -  Dalsgaard  \& Mullan  1994).  Also, we  have
coupled our pulsation code to  an stellar evolution code. We have done
it  in order  to be  able  to follow  the changes  in the  pulsational
spectrum of a given model during  its evolution with a number of steps
as large  as it would be  needed by specific studies.   In making such
computation we  selected an evolutionary  time step short  enough such
that eigenvalues and eigenfunctions corresponding to a given model are
used as  a initial approximation  to those of  the next one.   This is
very useful for a relaxation code like this and is intended to work as
it is the case of the Heneyey technique in stellar evolution.

The rest of the paper  is organized as follows: Section \ref{secc:deq}
is devoted to present the differential equations of linear, adiabatic,
nonradial pulsations  we have to solve.  In  Section \ref{secc:code} a
detailed  description of  the general  Newton -  Raphson  technique we
employ   for   computing  the   eigenmodes   is  presented.    Section
\ref{secc:polyt},  presents the  results  for the  case of  polytropic
spheres, and in Section \ref{secc:helium} we describe the work we have
done  with   a  realistic  helium   WD  model.  Finally,   in  Section
\ref{secc:conclu} summarize our results.

\section{The Differential Equations  of Nonradial Pulsations}
\label{secc:deq}

The differential equations that govern linear, nonradial pulsations of
spherical  stars in  the  adiabatic approximation  are  (Unno, et  al.
1989):

\begin{equation} \label{46}
{x\   {dy_{1}\over{dx}}}   =    {\Big(V_{g}   -   3\Big)}\   y_{1}   +
{\Bigg[{\ell(\ell+1)\over{C_{1}\   \omega^{2}}}  -  V_{g}\Bigg]}\   y_{2}  +
{V_{g}\ y_{3}},
\end{equation}

\begin{equation} \label{47}
{x\ {dy_{2}\over{dx}}} = {\Big(C_{1}\ \omega^{2} - A^{*}\Big)}\
y_{1} + {\Big(A^{*} - U + 1\Big)}\ y_{2} - A^{*}\ y_{3},
\end{equation}

\begin{equation} \label{48}
{x\ {dy_{3}\over{dx}}} = {\Big(1 - U\Big)}\ y_{3} + y_{4},
\end{equation}

\begin{equation} \label{49}
{x\ {dy_{4}\over{dx}}} = {U\ A^{*}\ y_{1}} + {U\ V_{g}\ y_{2}} +
{\bigg[\ell(\ell+1) - U\ V_{g}\bigg]}\ y_{3} - U\ y_{4}.
\end{equation}

\noindent $x$ is the independent variable defined as $x = r / R$ where
$r$  is the  radial  coordinate and  $R$  is the  stellar radius.  The
dependent variables are defined as

\begin{equation}
y_{1} = {\xi_{r}\over{r}},
\end{equation}

\begin{equation}
y_{2} = {1\over{g\ r}}\ {\Bigg({p^{'}\over{\rho}} +
\Phi^{'}\Bigg)},
\end{equation}

\begin{equation}
y_{3} = {1\over{g\ r}}\ \Phi^{'},
\end{equation}

and

\begin{equation}
y_{4} = {1\over{g}}\ {d\Phi^{'}\over{dr}},
\end{equation}

\begin{equation} \label{44}
\omega^{2} = {{\sigma^{2}\ R^{3}}\over{G\ M_{\star}}}.
\end{equation}

\noindent $\xi_{r}$  represents the  radial displacement of  the fluid
and  $p^{'}$  and  $\Phi^{'}$  are  the  Eulerian  variations  of  the
equilibrium values of the pressure $p$ and the gravitational potential
$\Phi$  respectively. $g$  is the  local acceleration  of  gravity and
$M_{\star}$  is the  stellar mass.  $\omega^{2}$ is  the dimensionless
square of the angular frequency of oscillation, $\sigma^{2}$.

Among the  coefficients of Eqs. (\ref{46}-\ref{49})  there appears the
harmonic  degree  $\ell$,  corresponding  to  the  spherical  harmonic
$Y_{\ell,m}(\theta,\phi)$  which describes  the angular  dependence of
the  oscillation pattern.  The  adimensional quantities  $V_{g}$, $U$,
$C_{1}$,  and $A^{*}$,  inherent to  the  non -  perturbed model,  are
defined as:

\begin{equation} \label{41}
V_{g} = \frac {V}{\Gamma_{1}}= -\ {1\over{\Gamma_{1}}}\ 
{d\ln p\over{d\ln r}} = {g\
r\over{c^{2}}} = {g\ r\ \rho\over{\Gamma_{1}\ p}},\label{eq:vg}
\end{equation}

\begin{equation} \label{42}
U = {d\ln M_r\over{d\ln r}} = {4\pi\ \rho\ r^{3}\over{M_r}},
\end{equation}

\begin{equation} \label{43}
C_{1} = {\Bigg({r\over{R}}\Bigg)^{3}}\ {M_{\star}\over{M_r}},
\end{equation}

\begin{equation} \label{45}
A^{*} = {{r\over{g}}\ N^{2}} = {r\ \Bigg({1\over{
\Gamma_{1}}}\ {d\ln p\over{dr}} - {d\ln \rho\over{dr}}\Bigg)}.\label{eq:astar}
\end{equation}

\noindent Here  $\Gamma_{1}$ is the first adiabatic  index, $c^{2}$ is
the square of the local velocity of sound, $M_r$ is the mass contained
in a (non - perturbed) sphere  of radius $r$, and $N^{2}$ is the Brunt
- V\"ais\"al\"a frequency.

Eqs.   (\ref{46}-\ref{49})  together   with   the  adequate   boundary
conditions conform the eigenvalue problem to be solved.

\section{The Numerical Code} \label{secc:code}

In  order to  write the  pulsation equations  in a  form  adequate for
numerical calculations we divide our non - perturbed model in a number
of  spherical, concentric  shells. This  is equivalent  to  divide the
dominion of the independent variable in $N$ mesh points ($N-1$ shells)
non -  necessarily equidistantly spaced  $x_{j}, j=1, \cdots,  N$.  In
our treatment  we define $x_{1}= 1$  as the surface and  $x_{N}= 0$ as
the central point of the model.

Now, we replace Eqs. (\ref{46}-\ref{49}) by difference equations.  The
system of equations may be rewritten as

\begin{equation}
{{dy_{i}}\over{dx}}= f_{i}(y_{1},y_{2},y_{3},y_{4}, \lambda),
\hspace{2cm} i=1, 2, 3, 4 \label{ec:gral},
\end{equation}

\noindent  where $\lambda=  \omega^{2}$.  In  finite  differences, Eq.
(\ref{ec:gral})  may  be  written  as\footnote{We have  also  employed
finite differences of the form

\begin{equation} \nonumber
{{[y_{i}]_{j+1} - [y_{i}]_{j}}\over{x_{j+1} - x_{j}}}=
\frac{1}{2}
\left( f_{i}\left([y_{1}, y_{2}, y_{3}, y_{4}]_{j};
\lambda\right)
+ \\ f_{i}\left([y_{1}, y_{2}, y_{3}, y_{4}]_{j+1};
\lambda\right)
\right);  i=1, 2, 3, 4; j=1, \cdots, N-1.
\end{equation}

\noindent  Applying such  form for  the difference  equations  we have
found  results  negligibly  different  from those  presented  in  this
paper.}

\begin{equation} \label{56}
{{[y_{i}]_{j+1} - [y_{i}]_{j}}\over{x_{j+1} - x_{j}}}=
f_{i}\left([y_{1}, y_{2}, y_{3},
y_{4}]_{j+\frac{1}{2}}; \lambda\right); i=1, 2, 3, 4; j=1, \cdots, N-1
\end{equation}

\noindent where, for any quantity $\psi$,

\begin{equation}
[\psi]_{j+\frac{1}{2}}=  {{[\psi]_{j} + [\psi]_{j+1}}\over{2}},
\end{equation}

\noindent being $[\psi]_{j}$  the value of the quantity  $\psi$ at the
meshpoint $j$. The outer boundary conditions are given by\footnote{The
first  two  of these  equations  are  the  so called  ``zero  boundary
condition'', which we shall employ  in the treatment of polytropes. In
the case  of realistic  WD models  we have replaced  the first  one by
Eq. (\ref{eq:buenaBC}) (see below).}  (Unno, et al. 1989)

\begin{equation}
\begin{tabular}{r}
$[y_{1}]_{1} - [y_{2}]_{1} + [y_{3}]_{1} = 0,$\\
$(\ell+1)\ [y_{3}]_{1} + [y_{4}]_{1}= 0,$\\
$[y_{1}]_{1}= 1.$
\end{tabular} \label{bc_outer}
\end{equation}

\noindent The  last equation  is the normalization  condition, usually
employed in previous works.  The central boundary conditions are given
by (Unno, et al. 1989)

\begin{equation}
\begin{tabular}{r}
$[y_{1}]_{N}\ {[C_{1}]_{N}\ \lambda\over{\ell}} - [y_{2}]_{N} = 0,$\\
$\ell\ [y_{3}]_{N} - [y_{4}]_{N} = 0.$
\end{tabular}
\end{equation}

In order to  solve this system  of difference equations  we shall
use  the   general Newton - Raphson technique  closely  following  
the formulation
presented in Kippenhahn,  et al. (1967)  for the case  of stellar
evolution. We  shall assume  we are  able to  get an  approximate
solution for the system and we want to improve it iteratively. In
the case that  this initial solution  is not far  from the actual
solution of the system, we  may expand the equations up  to first
order in the corrections for the values of the eigenfunctions  at
each meshpoint  and also  for the  eigenvalue (square of the
eigenfrequency). In this standard way  we get a linear system  of
equations in the corrections that must be solved.

In condensed way, the algebraic system of equations for the first
order corrections may be expressed as

\begin{equation} \label{73}
{\partial{B_{k}}\over{\partial{[y_{1}]_{1}}}}\ \delta [y_{1}]_{1} +
\cdots +
{\partial{B_{k}}\over{\partial{[y_{4}]_{1}}}}\ \delta [y_{4}]_{1} +
{\partial{B_{k}}\over{\partial{\lambda}}}\ \delta \lambda = {- B_{k}};\;
k = 1,2,3,
\end{equation}

\begin{equation} \label{74}
\begin{tabular}{l}
${\partial{G_{i}^{j}}\over{\partial{[y_{1}]_{j}  }}}\ \delta [y_{1}]_{j} + 
\cdots +
{\partial{G_{i}^{j}}\over{\partial{ [y_{4}]_{j}  }}}\ \delta [y_{4}]_{j}   +
{\partial{G_{i}^{j}}\over{\partial{ [y_{1}]_{j+1}}}}\ \delta [y_{1}]_{j+1} + 
\cdots +
{\partial{G_{i}^{j}}\over{\partial{ [y_{4}]_{j+1}}}}\ \delta [y_{4}]_{j+1} +
$\\ ${\partial{G_{i}^{j}}\over{\partial{\lambda}}}\ \delta \lambda =
{- G_{i}^{j}};\; i = 1,2,3,4; \;  j = 1,2,\ldots,N-1,$
\end{tabular}
\end{equation}

\begin{equation} \label{75}
{\partial{C_{m}}\over{\partial{[y_{1}]_{N}}}}\ \delta [y_{1}]_{N} + \cdots +
{\partial{C_{m}}\over{\partial{[y_{4}]_{N}}}}\ \delta [y_{4}]_{N} +
{\partial{C_{m}}\over{\partial{\lambda}}}\ \delta \lambda = {-
C_{m}};\;
m = 1,2,
\end{equation}

\noindent  where $\delta  [y_{i}]_{j}$  are small  corrections to  the
eigenfunction  $y_{i}$ at the  $j-$ mesh  point, and  $\delta \lambda$
stand  for  the  correction  to the  eigenvalue  $\lambda$.   $B_{k}$,
$G_{i}^{j}$, and  $C_{m}$ are the  values of the  difference equations
when applied  the solution to be iteratively  improved. Obviously, all
of them must be zero when evaluated with their exact solution.

Now, we have to invert a big matrix, that has non - zero elements only
in blocks located  on the diagonal and in the  last column (because of
the derivative with  respect to the eigenvalue).  Notice  that this is
an   important  difference   compared   with  the   case  of   stellar
evolution.  Thus,  in handling  the  big  matrix  we need  a  specific
algorithm to solve for the corrections.

Evaluating Eq.   (\ref{73}) for $k= 1,  2, 3$ and  Eq.  (\ref{74}) for
$i= 1,  2, 3,  4$ and $j=  1$, we  may write the  first block  of such
matrix as

\begin{equation}
\left[\begin{array}{ccccccc}
{\partial{B_{1}}\over{\partial{[y_{1}]_{1}}}} & \ldots & \ldots &
{\partial{B_{1}}\over{\partial{[y_{4}]_{1}}}} & 0 & 0 & 0\\
{\partial{B_{2}}\over{\partial{[y_{1}]_{1}}}} & \ldots & \ldots &
{\partial{B_{2}}\over{\partial{[y_{4}]_{1}}}} & 0 & 0 & 0\\
{\partial{B_{3}}\over{\partial{[y_{1}]_{1}}}} & \ldots & \ldots &
{\partial{B_{3}}\over{\partial{[y_{4}]_{1}}}} & 0 & 0 & 0\\
{\partial{G_{1}^{1}}\over{\partial{[y_{1}]_{1}}}} & \ldots &  \ldots &
{\partial{G_{1}^{1}}\over{\partial{[y_{4}]_{1}}}} &
{\partial{G_{1}^{1}}\over{\partial{[y_{1}]_{2}}}} &
{\partial{G_{1}^{1}}\over{\partial{[y_{2}]_{2}}}} &
{\partial{G_{1}^{1}}\over{\partial{[y_{3}]_{2}}}} \\
\vdots&  \ddots& & \vdots& \vdots& \vdots& \vdots\\
\vdots&  & \ddots& \vdots&  \vdots&  \vdots&  \vdots\\
{\partial{G_{4}^{1}}\over{\partial{[y_{1}]_{1}}}} & \ldots & \ldots &
{\partial{G_{4}^{1}}\over{\partial{[y_{4}]_{1}}}} &
{\partial{G_{4}^{1}}\over{\partial{[y_{1}]_{2}}}} &
{\partial{G_{4}^{1}}\over{\partial{[y_{2}]_{2}}}} &
{\partial{G_{4}^{1}}\over{\partial{[y_{3}]_{2}}}}
\end{array}\right]\ \cdot\
\left[\begin{array}{c}
\delta [y_{1}]_{1}\\
\delta [y_{2}]_{1}\\
\delta [y_{3}]_{1}\\
\delta [y_{4}]_{1}\\
\delta [y_{1}]_{2}\\
\delta [y_{2}]_{2}\\
\delta [y_{3}]_{2}
\end{array}\right] =
\left[\begin{array}{ccc}
0 & - {\partial{B_{1}}\over{\partial{\lambda}}} & - B_{1}\\
0 & - {\partial{B_{2}}\over{\partial{\lambda}}} & - B_{2}\\
0 & - {\partial{B_{3}}\over{\partial{\lambda}}} & - B_{3}\\
- {\partial{G_{1}^{1}}\over{\partial{[y_{4}]_{2}}}} &
- {\partial{G_{1}^{1}}\over{\partial{\lambda}}} &
- G_{1}^{1}\\
\vdots & \vdots & \vdots\\
\vdots &  \vdots & \vdots\\
- {\partial{G_{4}^{1}}\over{\partial{[y_{4}]_{2}}}} &
- {\partial{G_{4}^{1}}\over{\partial{\lambda}}} &
- G_{4}^{1}
\end{array}\right] \cdot\
\left[\begin{array}{c}
\delta [y_{4}]_{2}\\
\delta \lambda\\
1
\end{array}\right].
\end{equation}

\noindent Now, we define vectors $U_{s}, V_{s}, W_{s}, (s=1, \cdots, 4N -
5)$ such that we may write

\begin{equation} \label{78}
\left[\begin{array}{c}
\delta [y_{1}]_{1}\\
\delta [y_{2}]_{1}\\
\delta [y_{3}]_{1}\\
\delta [y_{4}]_{1}\\
\delta [y_{1}]_{2}\\
\delta [y_{2}]_{2}\\
\delta [y_{3}]_{2}
\end{array}\right] =
\left[\begin{array}{ccc}
U_{1} & V_{1} & W_{1}\\
U_{2} & V_{2} & W_{2}\\
\vdots & \vdots & \vdots\\
\vdots & \vdots & \vdots\\
\vdots & \vdots & \vdots\\
\vdots & \vdots & \vdots\\
U_{7} &  V_{7} &  W_{7}
\end{array}\right] \cdot\
\left[\begin{array}{c}
\delta [y_{4}]_{2}\\
\delta \lambda\\
1
\end{array}\right]
\end{equation}

\noindent for  this block.   The coefficients of  this vectors  may be
calculated easily solving

\begin{equation}
\left[\begin{array}{ccccccc}
{\partial{B_{1}}\over{\partial{[y_{1}]_{1}}}} & \ldots & \ldots &
{\partial{B_{1}}\over{\partial{[y_{4}]_{1}}}} & 0 & 0 & 0\\
{\partial{B_{2}}\over{\partial{[y_{1}]_{1}}}} & \ldots & \ldots &
{\partial{B_{2}}\over{\partial{[y_{4}]_{1}}}} & 0 & 0 & 0\\
{\partial{B_{3}}\over{\partial{[y_{1}]_{1}}}} & \ldots & \ldots &
{\partial{B_{3}}\over{\partial{[y_{4}]_{1}}}} & 0 & 0 & 0\\
{\partial{G_{1}^{1}}\over{\partial{[y_{1}]_{1}}}} & \ldots & \ldots &
{\partial{G_{1}^{1}}\over{\partial{[y_{4}]_{1}}}} &
{\partial{G_{1}^{1}}\over{\partial{[y_{1}]_{2}}}} &
{\partial{G_{1}^{1}}\over{\partial{[y_{2}]_{2}}}} &
{\partial{G_{1}^{1}}\over{\partial{[y_{3}]_{2}}}} \\
\vdots& \ddots& & \vdots& \vdots& \vdots& \vdots\\
\vdots& & \ddots& \vdots& \vdots& \vdots& \vdots\\
{\partial{G_{4}^{1}}\over{\partial{[y_{1}]_{1}}}} & \ldots & \ldots &
{\partial{G_{4}^{1}}\over{\partial{[y_{4}]_{1}}}} &
{\partial{G_{4}^{1}}\over{\partial{[y_{1}]_{2}}}} &
{\partial{G_{4}^{1}}\over{\partial{[y_{2}]_{2}}}} &
{\partial{G_{4}^{1}}\over{\partial{[y_{3}]_{2}}}}
\end{array}\right]\ \cdot\
\left[\begin{array}{ccc}
U_{1} & V_{1} & W_{1}\\
U_{2} & V_{2} & W_{2}\\
\vdots & \vdots & \vdots\\
\vdots & \vdots & \vdots\\
\vdots & \vdots & \vdots\\
\vdots & \vdots & \vdots\\
U_{7} & V_{7} & W_{7}
\end{array}\right] =
\left[\begin{array}{ccc}
0 & - {\partial{B_{1}}\over{\partial{\lambda}}} & - B_{1}\\
0 & - {\partial{B_{2}}\over{\partial{\lambda}}} & - B_{2}\\
0 & - {\partial{B_{3}}\over{\partial{\lambda}}} & - B_{3}\\
- {\partial{G_{1}^{1}}\over{\partial{[y_{4}]_{2}}}} &
- {\partial{G_{1}^{1}}\over{\partial{\lambda}}} &
- G_{1}^{1}\\
\vdots & \vdots & \vdots\\
\vdots & \vdots & \vdots\\
- {\partial{G_{4}^{1}}\over{\partial{[y_{4}]_{2}}}} &
- {\partial{G_{4}^{1}}\over{\partial{\lambda}}} &
- G_{4}^{1}
\end{array}\right].
\end{equation}

\noindent From  Eq. (\ref{74}), and  after short manipulation,  we may
write an arbitrary block of the big matrix (except the central) in the
form

\begin{equation} \label{90}
\left[\begin{array}{c}
\delta [y_{4}]_{j}\\
\delta [y_{1}]_{j+1}\\
\delta [y_{2}]_{j+1}\\
\delta [y_{3}]_{j+1}
\end{array}\right] =
\left[\begin{array}{ccc}
U_{4j}   & V_{4j}   & W_{4j}\\
U_{4j+1} & V_{4j+1} & W_{4j+1}\\
U_{4j+2} & V_{4j+2} & W_{4j+2}\\
U_{4j+3} & V_{4j+3} & W_{4j+3}
\end{array}\right] \cdot\
\left[\begin{array}{c}
\delta [y_{4}]_{j+1}\\
\delta \lambda\\
1\\
\end{array}\right],
\end{equation}

\noindent where $j = 2,\ldots, N-2$.  Now, the coefficients of vectors
$U_{s}, V_{s}, W_{s}$ may be evaluated solving the system

\begin{equation}
\left[\begin{array}{cccc}
\alpha_{1}^{j} &
{\partial{G_{1}^{j}}\over{\partial{[y_{1}]_{j+1}}}} &
{\partial{G_{1}^{j}}\over{\partial{[y_{2}]_{j+1}}}} &
{\partial{G_{1}^{j}}\over{\partial{[y_{3}]_{j+1}}}}\\
\alpha_{2}^{j} &
{\partial{G_{2}^{j}}\over{\partial{[y_{1}]_{j+1}}}} &
{\partial{G_{2}^{j}}\over{\partial{[y_{2}]_{j+1}}}} &
{\partial{G_{2}^{j}}\over{\partial{[y_{3}]_{j+1}}}}\\
\alpha_{3}^{j} &
{\partial{G_{3}^{j}}\over{\partial{[y_{1}]_{j+1}}}} &
{\partial{G_{3}^{j}}\over{\partial{[y_{2}]_{j+1}}}} &
{\partial{G_{3}^{j}}\over{\partial{[y_{3}]_{j+1}}}}\\
\alpha_{4}^{j} &
{\partial{G_{4}^{j}}\over{\partial{[y_{1}]_{j+1}}}} &
{\partial{G_{4}^{j}}\over{\partial{[y_{2}]_{j+1}}}} &
{\partial{G_{4}^{j}}\over{\partial{[y_{3}]_{j+1}}}}
\end{array}\right] \cdot\
\left[\begin{array}{ccc}
U_{4j}   & V_{4j}   & W_{4j}\\
U_{4j+1} & V_{4j+1} & W_{4j+1}\\
U_{4j+2} & V_{4j+2} & W_{4j+2}\\
U_{4j+3} & V_{4j+3} & W_{4j+3}
\end{array}\right] =
\left[\begin{array}{ccc}
- {\partial{G_{1}^{j}}\over{\partial{[y_{4}]_{j+1}}}} &
- \beta_{1}^{j} &
- \gamma_{1}^{j}\\
- {\partial{G_{2}^{j}}\over{\partial{[y_{4}]_{j+1}}}} &
- \beta_{2}^{j} &
- \gamma_{2}^{j}\\
- {\partial{G_{3}^{j}}\over{\partial{[y_{4}]_{j+1}}}} &
- \beta_{3}^{j} &
- \gamma_{3}^{j}\\
- {\partial{G_{4}^{j}}\over{\partial{[y_{4}]_{j+1}}}} &
- \beta_{4}^{j} &
- \gamma_{4}^{j}\\
\end{array}\right],
\end{equation}

\noindent where quantities $\alpha_{i}^{j}, \beta_{i}^{j}, \gamma_{i}^{j}$
are defined as

\begin{equation}
\begin{tabular}{lclclclcl}
$\alpha_{i}^{j}$ &=&
${\partial{G_{i}^{j}}\over{\partial{[y_{4}]_{j}}}}$ &+&
$U_{4j-3}\ {\partial{G_{i}^{j}}\over{\partial{[y_{1}]_{j}}}}$ &+&
$U_{4j-2}\ {\partial{G_{i}^{j}}\over{\partial{[y_{2}]_{j}}}}$ &+&
$U_{4j-1}\ {\partial{G_{i}^{j}}\over{\partial{[y_{3}]_{j}}}},$\\
$$ & $$ & $$ & $$ & $$ & $$ & $$ & $$ & $$\\
$\beta_{i}^{j}$ &=&
${\partial{G_{i}^{j}}\over{\partial{\lambda}}}$ &+&
$V_{4j-3}\ {\partial{G_{i}^{j}}\over{\partial{[y_{1}]_{j}}}}$ &+&
$V_{4j-2}\ {\partial{G_{i}^{j}}\over{\partial{[y_{2}]_{j}}}}$ &+&
$V_{4j-1}\ {\partial{G_{i}^{j}}\over{\partial{[y_{3}]_{j}}}},$\\
$$ & $$ & $$ & $$ & $$ & $$ & $$ & $$ & $$\\
$\gamma_{i}^{j}$ &=& $G_{i}^{j}$ &+&
$W_{4j-3}\ {\partial{G_{i}^{j}}\over{\partial{[y_{1}]_{j}}}}$ &+&
$W_{4j-2}\ {\partial{G_{i}^{j}}\over{\partial{[y_{2}]_{j}}}}$ &+&
$W_{4j-1}\ {\partial{G_{i}^{j}}\over{\partial{[y_{3}]_{j}}}}.$\\
$$ & $$ & $$ & $$ & $$ & $$ & $$ & $$ & $$\\
\end{tabular} \label{eq:alphabetagamma}
\end{equation}

\noindent with $i= 1, 2, 3, 4$.

Finally, in the central point of the model, we have $j= N - 1$ in
Eq. (\ref{74}) and $m=  1, 2$ in (\ref{75}).  It is easy to  show
that the last block of the big matrix may be written as

\begin{equation} \label{96}
\left[\begin{array}{cccccc}
\alpha_{1}^{N-1} &
{\partial{G_{1}^{N-1}}\over{\partial{[y_{1}]_{N}}}} &
{\partial{G_{1}^{N-1}}\over{\partial{[y_{2}]_{N}}}} &
{\partial{G_{1}^{N-1}}\over{\partial{[y_{3}]_{N}}}} &
{\partial{G_{1}^{N-1}}\over{\partial{[y_{4}]_{N}}}} &
\beta_{1}^{N-1}\\
\alpha_{2}^{N-1} &
{\partial{G_{2}^{N-1}}\over{\partial{[y_{1}]_{N}}}} &
{\partial{G_{2}^{N-1}}\over{\partial{[y_{2}]_{N}}}} &
{\partial{G_{2}^{N-1}}\over{\partial{[y_{3}]_{N}}}} &
{\partial{G_{2}^{N-1}}\over{\partial{[y_{4}]_{N}}}} &
\beta_{2}^{N-1}\\
\alpha_{3}^{N-1} &
{\partial{G_{3}^{N-1}}\over{\partial{[y_{1}]_{N}}}} &
{\partial{G_{3}^{N-1}}\over{\partial{[y_{2}]_{N}}}} &
{\partial{G_{3}^{N-1}}\over{\partial{[y_{3}]_{N}}}} &
{\partial{G_{3}^{N-1}}\over{\partial{[y_{4}]_{N}}}} &
\beta_{3}^{N-1}\\
\alpha_{4}^{N-1} &
{\partial{G_{4}^{N-1}}\over{\partial{[y_{1}]_{N}}}} &
{\partial{G_{4}^{N-1}}\over{\partial{[y_{2}]_{N}}}} &
{\partial{G_{4}^{N-1}}\over{\partial{[y_{3}]_{N}}}} &
{\partial{G_{4}^{N-1}}\over{\partial{[y_{4}]_{N}}}} &
\beta_{4}^{N-1}\\
0 &
{\partial{C_{1}}\over{\partial{[y_{1}]_{N}}}} &
{\partial{C_{1}}\over{\partial{[y_{2}]_{N}}}} &
{\partial{C_{1}}\over{\partial{[y_{3}]_{N}}}} &
{\partial{C_{1}}\over{\partial{[y_{4}]_{N}}}} &
{\partial{C_{1}}\over{\partial{\lambda}}}\\
0 &
{\partial{C_{2}}\over{\partial{[y_{1}]_{N}}}} &
{\partial{C_{2}}\over{\partial{[y_{2}]_{N}}}} &
{\partial{C_{2}}\over{\partial{[y_{3}]_{N}}}} &
{\partial{C_{2}}\over{\partial{[y_{4}]_{N}}}} &
{\partial{C_{2}}\over{\partial{\lambda}}}\\
\end{array}\right]\ \cdot
\left[\begin{array}{c}
\delta [y_{4}]_{N-1}\\
\delta [y_{1}]_{N}\\
\delta [y_{2}]_{N}\\
\delta [y_{3}]_{N}\\
\delta [y_{4}]_{N}\\
\delta \lambda\\
\end{array}\right]\ =
\left[\begin{array}{c}
-\gamma_{1}^{N-1}\\
-\gamma_{2}^{N-1}\\
-\gamma_{3}^{N-1}\\
-\gamma_{4}^{N-1}\\
-C_{1}\\
-C_{2}\\
\end{array}\right].
\end{equation}

\noindent   In   this    case   the   quantities   $\alpha_{i}^{N-1}$,
$\beta_{i}^{N-1}$,   $\gamma_{i}^{N-1}$   are   evaluated  from   Eqs.
(\ref{eq:alphabetagamma}) at $j= N-1$.

We  note that  Eq. (\ref{96})  can be  solved in  order to  obtain the
corrections for  eigenfunctions $y_{1}$, $y_{2}$,  $y_{3}$ and $y_{4}$
in the central point of  the object, and for the eigenvalue $\lambda$.
Also, it  must be noted  that the correction in  eigenfunction $y_{4}$
pertaining  to the  immediate outer  mesh point  of grid  is obtained.
This correction, namely  $\delta [y_{4}]_{N-1}$, serve as ``coupling''
between the points  $N$ and $N-1$. In fact, we  may use Eq. (\ref{90})
with $j= N-2$ in order to  obtain the rest of corrections belonging to
the eigenfunctions at  the mesh point $N-1$.  Next,  the employment of
this   procedure  for   successive   downwards  values   of  $j$   (in
Eq.  \ref{90}) using  $\delta [y_{4}]_{j+1}$  as coupling  between the
quantities  belonging  to  consecutive  meshpoints,  as  well  as  Eq.
(\ref{78}),  leads to  find  the corrections  for  the eigenvalue  and
eigenfunctions for the complete  model.  These corrections are applied
to the initial  solution and the algorithm is  employed iteratively up
to the stage at which all  the (absolute value) of the corrections are
smaller  than  some previously  adopted  value.   At  this point,  the
complete    set    of     difference    equations    that    represent
Eqs. (\ref{46}-\ref{49}) has been solved.

In order to  search for the first approximation  to the eigenfunctions
and  the eigenvalue  of a  mode,  we have  applied the  method of  the
discriminant  presented  in Unno  et  al.  (1989).  Specifically,  the
expression   adopted   reads    $D(\omega)=   (\ell+1)\   [y_{3}]_{1}   +
[y_{4}]_{1}$.  Notice that  $D(\omega)=0$ corresponds  exactly  to the
second outer  boundary condition given in  Eqs.  (\ref{bc_outer}).  We
refer the reader to that book for more details.

Obviously, for  applying this  technique we need  to define a  grid of
meshpoints.  In  getting an appropriate distribution  of meshpoints we
have  employed a  simple  recipe.   As a  first  approximation to  the
solution   of  the   oscillation  equations   we  have   computed  the
eigenfunctions  and   eigenvalue  taking  the  points   at  which  the
equilibrium   model   is   defined.    After  convergence,   we   have
redistributed the grid asking  for the same criterium usually employed
in  stellar evolution  (Kippenhahn, et  al. 1967):  that  the relative
variation  of each  eigenfunction inside  a  zone must  be below  some
prescribed  limit (if  the eigenfunctions  are below  some  other very
small,  prescribed value  we have  asked for  absolute  differences in
place  of relative  ones). If  necessary,  our program  add or  remove
meshpoints,  using   spline  interpolation  over   quantities  of  non -
perturbed  model, and  lineal interpolation  in  eigenfunctions, since
these will be improved after in the  general Newton - Raphson  stage.

\section{Application  to  Polytropic   Spheres}
\label{secc:polyt}

As a first test  for our code, we have applied it  to the well - known
case of nonradial pulsations of polytropic spheres.  Polytropes are
particularly adequate for the purpose of testing, because in this case
we separate the  inaccuracies of the non -  perturbed model from those
inherent to  the nonradial pulsation  code. In particular  we have
computed the p- (pressure), f- (fundamental) and g- (gravity) modes of
polytropes with indices $n= 1.5, 2,  2.5, 3, 3.5$ and $4$ for the case
of $\ell= 2, 3$ and $4$.

Because in this case $V_{g}$ and $A^{*}$ are divergent at surface (see
Eqs. \ref{eq:vg} and \ref{eq:astar}), we have integrated the structure
by means of an accurate Runge  - Kutta technique (Press, et al.  1992)
and most of the meshpoints  were located near surface. Here we assumed
$\Gamma_{1}= \frac{5}{3}$.

To compare our  eigenvalues with those available in  the literature we
shall  employ the  work of  Christensen -  Dalsgaard \&  Mullan (1994)
which presents  accurate eigenfrequency  tabulations for a  variety of
polytropic  spheres.   For  such  purpose  we  have  plotted  in  Fig.
\ref{fig:chd}            the            relative            difference
$|\omega^{2}-\omega_{Ch-D}^{2}|/\omega_{Ch-D}^{2}$  for  the cases  of
p-modes for $n= 1.5,  3$ and $\ell= 2, 3$, g-modes for  $n= 3$ and $\ell= 2,
3, 4$,  and also for p-modes for  $n=4$ and $\ell= 2,  3$. The comparison
indicates  that  the  higher  radial  order the  larger  the  relative
differences are,  without any dependency  on $\ell$ value.   The absolute
value of relative differences found in the eigenvalues are $\lesssim 4
\times  10^{-4}$,  except  in  the  $n=  3$  g-modes,  for  which  the
difference is $\lesssim 2 \times 10^{-3}$.  As Christensen - Dalsgaard
\& Mullan (1994)  state that their computations are  accurate to $\sim
10^{-8}$ we conclude that our  code is able to compute the eigenvalues
of polytropes with  a precision of $\sim 10^{-3}$  which is enough for
our purposes.

\section{Application to Realistic Models: Nonradial g-Modes in Helium 
White Dwarfs} \label{secc:helium}

With the  aim of investigating  the behaviour of  the code when  it is
applied  to  realistic  models,  in  this section  we  shall  consider
nonradial  g-modes in  evolutionary models  of low  mass,  pure helium
WDs. This  election is due mainly  because of two  reasons.  First, at
present we have available a detailed and updated code we have employed
in the calculation of the  evolution of such stars (see, e.g., Althaus
\& Benvenuto 1997; Benvenuto \& Althaus 1998).  Such evolutionary code
is fully described in  the above-cited papers. Second, the computation
of nonradial modes in pure helium  WDs provide us an important way for
testing  the pulsation  code  when applied  to  realistic models.   At
present, to our  notice there is no study  available in the literature
on  the pulsation  properties  of these  objects.   In evaluating  our
results we shall pay special  attention to the asymptotic behaviour of
periods  of  oscillation,  comparing  in  particular  the  spacing  of
adjacent  periods  (for  the  same  $\ell$)  with  that  predicted  by
asymptotic  theory  of  Tassoul   (1980)  in  stars  with  homogeneous
composition. This theory has been  employed as check for the nonradial
spectrum of  polytropes (Mullan  \& Ulrich 1988;  Mullan 1989)  in the
Cowling approximation.

\subsection{Outline of Pulsation Calculations}

One  possible  way  for  computing nonradial  oscillations  modes  in
evolutionary  models  is  to  first  compute such  structures  with  a
evolutionary  code,  and then  to  choose  a  subset of  these  models
(usually belonging  to a  predetermined interval in  $T_{eff}$, called
hereafter $T_{eff}$-strip) for  pulsation analysis. This procedure has
been employed in most of the studies of adiabatic pulsations in WDs to
date  (see, e.g.,  TFW, Bradley,  Winget \&  Wood 1993,  Bradley 1996,
Brassard et al. 1992a, 1992b).

Here, we present an alternative  way for computing nonradial modes in
evolutionary sequences.   The basic idea  is very direct: if  the time
step  in the evolutionary  sequence is  short enough,  eigenvalues and
eigenfunctions corresponding to a given model should strongly resemble
those  corresponding to  the previous  one.   Then, if  we couple  the
pulsation code to  the evolutionary code, it is  feasible to avoid the
scan in  $\omega^{2}$ for each model:  the search of  modes is carried
out only for the first model inside the relevant $T_{eff}$-strip.  The
modes are  computed and stored and  serve as a trial  solution for the
next model and  so on. In this way it is  possible economize CPU time,
and (more  importantly) it  is possible to  follow the changes  in the
pulsational spectrum due to the  evolution of the stellar structure in
a continuous fashion.

\subsection{Details of Calculations}

Let us briefly describe how  our pulsation and evolutionary codes work
together.   Firstly,  $T_{eff}$-strip  is   chosen,  as  well  as  the
frequency  window to be  scanned. The  evolutionary code  computes the
model up  to the  moment when the  model reaches  the hot edge  of the
$T_{eff}$-strip.   Then,  the  program  calls  the  pulsation  routine
beginning  the   scan  for   eigenfrequencies,  as  described   in  \S
\ref{secc:code}.  When  a  mode   is  found,  the  code  generates  an
approximate  solution for  $y_{i}; i=1,  \cdots, 4$  and $\omega^{2}$,
which is improved iteratively.  Then,  such solution is tested and, if
necessary,   meshpoints   are   redistributed   as  outlined   in   \S
\ref{secc:code},  and iterated to  convergence.  Before  improving the
solution, the  eigenmodes have been stored  on the original  grid in a
common block, with  the aim of being employed  later as an approximate
solution  for  the next  stellar  model  of  the sequence.   When  the
computation of the mode has been finished, the code begins to look for
eigenmodes  again,  and  the  process  is  repeated  until  the  given
frequency  window  is  fully  covered.  Thus,  we  have  finished  the
computation  of all  eigenmodes of  the first  model belonging  to the
$T_{eff}$-strip and the  structure of each one is  now in the computer
memory. Then,  the evolutionary code generates the  next stellar model
and the code  calls pulsation routine again, but  in this occasion the
search for modes  is skipped.  Instead of this,  stored eigenmodes are
taken as input  to the  general Newton - Raphson  scheme to approximate
the modes of this
subsequent  stellar model.  Then,  the solution  is  iterated, and  so
on. The whole procedure is automatically repeated for all evolutionary
models  inside the  $T_{eff}$-strip. When  the star  gets  outside the
$T_{eff}$-strip, computation is finished.

In terms of CPU, a large fraction  of the running time is spent in the
search  for modes,  but this  is executed  only once.  The  process of
relaxation and improvement of the solution is faster, and the building
of each stellar model is almost instantaneous.

\subsection{Helium White Dwarfs Models}

As mentioned before, we have  choose models of helium WDs for checking
the   efficiency    our   pulsation   code    in   realistic   stellar
configurations. To  be specific, we computed nonradial  g-modes for a
helium  WD model with  mass $M_{*}=  0.3 M_{\odot}$,  in the  range of
$8000K \leq  T_{eff} \leq 25000K$. The complete  sequence comprise 216
models. Convection,  present in the outermost layers,  is treated with
the ML3 version of Mixing Length Theory.

\subsection{Results}

Since  one  of  our  interests  here is  to  evaluate  the  asymptotic
behaviour of eigenperiods, we have computed nonradial g-modes for $\ell=
1, 2$ and $3$ with radial order from $k= 1$ to $56$. The resulting set
of modes is adequate for our purpose, since it covers a broad range in
periods, enough  for enabling the  comparison with the  predictions of
the asymptotic theory of Tassoul (1980). Here we employ the mechanical
external boundary condition given in Unno, et al. (1989)

\begin{equation}
y_{1} \bigg[ 1 + \bigg( \frac{\ell(\ell+1)}{\omega^2}
- 4 - \omega^2 \bigg) \frac{1}{V} \bigg] - y_{2} +
y_{3} \bigg[ 1 + \bigg( \frac{\ell(\ell+1)}{\omega^2} - \ell- 1 \bigg)
\frac{1}{V} \bigg] = 0,
\label{eq:buenaBC}\end{equation}

\noindent which replaces the first of Eqs. (\ref{bc_outer}).

For  each computed  mode, the  quantities of  interest are  the period
$P_{k}$, eigenfunctions  $y_{i}; i=1, \cdots,4$;  kinetic energy K.E.,
and  first  order rotation  splitting  coefficient, $C_{\ell,k}$.   In
addition,  we compute  the  variational period,  $P_{k}^{v}$, and  the
weight function,  WF, in the form  given by Kawaler,  Hansen \& Winget
(1985). Also,  for each non-perturbed  model we obtain  the asymptotic
spacing of  periods (for  the same $\ell$),  $\Delta P_{A}$,  given by
(TFW, Tassoul 1980)

\begin{equation}
\Delta P_{A} = \frac{P_{0}}{\sqrt{\ell(\ell+1)}},
\label{eq:asympt}
\end{equation}

\noindent where $P_{0}$ is defined as

\begin{equation}
P_{0} = 2 \pi^2 \bigg[\int_{0}^{1} \frac{N}{x} dx \bigg]^{-1}.
\end{equation}

As in  TFW, for computing $P_{0}$  we have ignored the  presence of an
OCZ, integrating from  the center to the surface  of model but setting
$N= 0$ in  the OCZ where $N^{2} < 0$.  In  this approximation, we have
overestimated the value of $\Delta  P_{A}$, but since the thickness of
the  OCZ (in  the radial  coordinate) is  very small,  this is  a good
approximation.

The  results obtained  for  $\ell=  1, 2,  3$  are qualitatively  very
similar.   Although pulsations  have not  yet been  observed  in these
stars, in  other pulsating  WDs the modes  with $\ell=1$  dominate the
observations, thus we shall concentrate in $\ell= 1$.

In Fig. \ref{fig:he_y1} we  have plotted the eigenfunction $y_{1}$ for
the modes  g$_{1}$, $\cdots$, g$_{5}$ with  $\ell= 1, 2$,  for a model
with  $T_{eff}= 11900$ K.   From Fig.   \ref{fig:he_y1} it  is evident
that $y_{1}$ has large amplitudes in the whole star, especially in the
core.  This feature is strongly  emphasized when we inspect modes with
increasing $k$.  This  is a remarkable difference in  the behaviour of
$y_{1}$ comparing with the case of  oscillating DA and DB WDs. In such
WDs $y_{1}$ has a lower  central amplitude (for modes with $\ell=2$ in
a $0.6 M_{\odot}$ DA WD model see, e.g., Fig.  14a and 16a of Brassard
et al.  1992b).

Fig.  \ref{fig:he_wf} displays the normalized weight function (WF) for
g$_{1}$,$\cdots$,g$_{5}$ ($\ell= 1$) of the same model.  WF provides a
measure of the relative contribution from different zones in the model
to  the period  formation. The  plot indicates  that  period formation
happens mainly in external region of star, where WF is large.  However
there are  important contributions from central  locations ($x \gtrsim
0.2$), which is somewhat different to the WF corresponding to the case
of DA and DB WDs, for  which WF displays appreciate values only in the
envelope (in the ``normal'' modes; see  Figs. 15 and 17 of Brassard et
al. 1992b).

Let us  discuss our results  concerning periods ($P_{k}$)  and kinetic
energies (K.E.).  The periods here computed show the expected trend in
g-modes, with  values increasing with  the order of modes  $k$.  Since
our calculations  covers a wide $T_{eff}$-strip, it  is possible infer
the changes of periods during WD cooling.

It is worth mentioning that we have computed the Brunt - V\"ais\"al\"a
frequency as  in TFW  (the Ledoux term  $B$ is  zero for the  whole of
model,  since  it is  chemically  homogeneous).   We  have found  that
$N^{2}$  decreases as the  model cools  down, a  feature that  is very
pronounced in the core. It  is due mainly to the increasing degeneracy
in  that region, decoupling  more and  more pressure  from temperature
($\chi_{T}$    drops).    As    consequence    eigenperiods   increase
monotonically when  $T_{eff}$ drops below  $\sim 20000$ K,  as clearly
shown in Fig.  \ref{fig:he_per}A.

Next, we  consider the behaviour of  the kinetic energy  (K.E.) of the
modes.  We  have found that,  as expected in a  chemically homogeneous
star, K.E.   is a smooth function of  $k$ and thus of  the period.  In
Fig.  \ref{fig:he_ener}  we show $\log$(K.E.)  vs.   $T_{eff}$ for the
models included  in Fig.  \ref{fig:he_per}.   For hotter models  it is
clear  that the  more  energetic  modes are  those  with lower  order,
because  they penetrate  deep  in  the star,  where  density is  high.
However, when $T_{eff}$ drops below  $\sim 20000$ K the energy of high
order modes  (which are mainly  concentrated in the outer  envelope of
star) strongly increase. The  explanation of this effect (see Brassard
et al.  1992b in context of a  DA WD star) resides in the fact that at
such $T_{eff}$ value the OCZ  of the object suddenly gets thicker (see
the dotted  line in Fig. 17  of Althaus \& Benvenuto  1997).  The high
order  modes  that  oscillates  in  the envelope  feel  gradually  the
presence of  convection as  the star cools  down.  Since that  in OCZs
g-modes  become  evanescent,  such  modes  are forced  to  get  larger
amplitudes somewhat below  the bottom of the OCZ  where the density is
larger.   Since K.E.   is proportional  to the  integral of  square of
displacement, weighted  by $\rho$,  these modes oscillate  with larger
energies. In  sharp contrast to the  behaviour of high  $k$ modes, low
order modes are rather insensitive to the thickening of the OCZ.

Now let us discuss the period  spacing of the modes.  To our knowledge
there is  no study  available in the  literature devoted  to nonradial
pulsations of helium WDs.  This situation does not allow us to perform
a direct comparison  of our results.  Nevertheless we  can have a good
idea related  to the reliability  of our pulsation code  examining the
period  spacing of  consecutive  modes of  the  same $\ell$.   Tassoul
(1980) predicted that the asymptotic  period spacing for g-modes for a
chemically  homogeneous object  in the  adiabatic approximations  is a
constant  value $\Delta  P_{A}$ (see  Eq.  \ref{eq:asympt}).   In Fig.
\ref{fig:DeltaP}  we show  the forward  period spacing  $\Delta P_{k}=
P_{k+1}  - P_{k}$  vs.  radial  order $k$  ($\ell=1$) for  models with
$T_{eff}= 13200, 11800$ and $9400$  K. As reference we also include in
this figure the corresponding  asymptotic spacing predicted by Tassoul
(1980) treatment.  From  Fig.  \ref{fig:DeltaP}  it is clear  that the
trend of the numerical solution is the correct one.

In order  to get an  estimation of the  accuracy of our  treatment, we
have  calculated the variational  eigenperiods ($P^{v}_{k}$)  for each
mode  in the  formalism presented  by  Kawaler, et  al.  (1985).   The
comparison of variational and numerical eigenperiods for the values of
$\ell$  here considered  gives a  difference between  these treatments
which is lower than $\approx 1 \%$ (see Fig. \ref{fig:variational} for
the case of $\ell=1$).

By  the way,  let us  comment on  that we  have also  tested  our code
running  it  on the  same  models  for  which oscillation  modes  were
previously computed.  Specifically we  have employed for such test two
carbon - oxygen  DA WD models of 0.5  $M_{\odot}$ and 0.85 $M_{\odot}$
whose structure was  computed with the WDEC evolutionary  code and its
vibrational  properties were  previously analyzed  (Bradley  1996). In
this  case $M_{H}/M_{*}=10^{-4}$; $M_{He}/M_{*}=10^{-2}$  and $T_{eff}
\approx  12500$K. This  is  a  valuable test  for  our pulsation  code
because we could confront our  results with those of other researchers
working  on detailed  models. Also  the  fact that  these models  were
computed  with   the  same  evolutionary  code   is  valuable  because
differences in  the resulting  periods are due  only to  the pulsation
code  and not to  differences in  e.g.  between  the treatment  of the
equation of state in our  evolutionary code and in WDEC. Considering a
large amount of modes, the differences between the previously computed
periods and those we have found has been below $\approx 0.1 \%$. Thus,
we  judge  that our  code  produce  results  accurate enough  for  our
purposes.  This comparison has been  made possible by Paul Bradley who
provided us with the DA WD models and the periods of many its modes.

\section{Summary and Conclusions} \label{secc:conclu}

In  this  paper  we  have   presented  a  general  code  intended  for
calculating   linear,    adiabatic,   nonradial   eigenfunctions   and
eigenfrequencies corresponding to spherically symmetric stars by means
of a finite difference scheme. This is a very simple strategy based on
a  slight modification  of the  general  Newton -  Raphson method  for
computing  stellar  evolution  presented  by  Kippenhahn,  Weigert  \&
Hofmeister (1967) because of the  different structure of the matrix to
be inverted in finding the iterative corrections.

As a first test to this code  we have applied it to the simplest case:
polytropic  spheres. As there  exist very  precise tabulations  of the
eigenfrequencies for such configurations, we have been able to perform
a quantitative  comparison of our  results with the best  available in
the literature (Christensen - Dalsgaard \& Mullan 1994). We have found
that our code  is able to compute polytropic  eigenfrequencies with an
accuracy of $10^{-3}$ or much better for low $k$ values.

Then,  we  have  coupled   this  nonradial  oscillation  code  to  our
evolutionary code and  applied it to the computation  of the pulsation
spectrum of pure, low mass  helium white dwarfs (WDs). Specifically we
considered the case of an object  with a mass of 0.3 $M_{\odot}$ for a
wide  range of  effective temperatures  ($T_{eff}$-strip).   In making
such  computation we have  employed a  new method.   Taking a  short -
enough  time  step  in  the  evolutionary  sequence,  eigenvalues  and
eigenfunctions corresponding to a  given model strongly resemble those
corresponding to  the previous  one.  Then, we  scan for modes  in the
first model  inside the $T_{eff}$-strip.  Eigenmodes  are computed and
stored and serve as a trial solution  for the next model and so on. In
this way it  is possible to follow the changes  in the the pulsational
spectrum due to the evolution of the stellar structure in a continuous
and fairly detailed fashion (see Fig. \ref{fig:he_per}).

Regarding  the computation  of g-modes  for the  0.3  $M_{\odot}$ pure
helium WD  model we have presented eigenfunctions  and weight function
(WF) for  low order  $k$ modes, and  periods, period  spacing, kinetic
energy (K.E.) and variational periods for a wide range of modes. These
quantities were computed  for the case of $\ell= 1, 2$  and $3$ but in
Figs. \ref{fig:he_wf} - \ref{fig:variational} we have presented only
the case $\ell=1$ for brevity. To  our notice this is the first effort
in studying the pulsation properties of helium WDs.

Our  experience with  realistic helium  WD models  indicates  that our
numerical  code  for pulsations  work  nicely  and  in particular  the
solutions  we have  found have  an asymptotic  period spacing  in good
agreement  with the predictions  of the  analytical theory  of Tassoul
(1980). This, and the fact that comparison with more complex WD models
with  layers  of  different  internal  composition show  us  that  our
numerical  tool for computing  eigenmodes work  adequately, especially
regarding  the   coupling  we  have  performed  between   it  and  our
evolutionary code. We  plan in the near future to  apply this new code
to the study of pulsation of WDs in general.

\acknowledgments  The authors  would like  to warmly  acknowledge Paul
Bradley  for his  kindness in  providing us  with numerical  models of
carbon -  oxygen DA WDs  together with his pulsation  calculations and
also for the time he spent in doing so. This allowed us to perform key
a test to our  code, such that it made possible to  us to be confident
with the results the code produces.  We also want to thank our referee
D.  Winget  for suggestions  that allowed us  to improve  the original
version of this paper.




\begin{figure} \caption{The absolute value of relative difference
of the  square of the eigenfrequency  for the case  of polytropes with
$n=1.5, 3$ and $4$, for  $\ell= 2$ (filled circles), $\ell= 3$ (filled
squares)  and $\ell=  4$  (filled triangles)  compared  with the  very
accurate calculations  of Christensen  - Dalsgaard and  Mullan (1994).
The points corresponding to modes of the same degree are connected for
clarity.  Notice  that differences between those  sets of computations
are larger the larger is the order  $k$ of the mode (i.e., in the case
of strongly oscillating eigenfunctions).}
\label{fig:chd} \end{figure}


\begin{figure} \caption{{\bf A} The eigenfunction $y_{1}$ for the modes 
g$_{1}$, $\cdots$,  g$_{5}$ with $\ell=  1$, corresponding to  the 0.3
$M_{\odot}$ pure  helium WD model with $T_{eff}=  11900$K.  Notice the
large amplitude of such modes in the stellar core. {\bf B} Same as {\bf
A} but with $\ell=2$.}
\label{fig:he_y1}
\end{figure}


\begin{figure} \caption{The normalized
weight  function  WF  corresponding  to  the same  modes  included  in
Fig. \ref{fig:he_y1}{\bf A}.}\label{fig:he_wf}
\end{figure}


\begin{figure} \caption{{\bf A} Periods of dipolar modes ($\ell=1$) with
radial order from $k= 1$ to  $k= 56$ and {\bf B} the asymptotic period
spacing  $\Delta  P_{A}$  predicted  by  Tassoul (1980)  theory  as  a
function  of the effective  temperature.  Notice  that the  periods of
high  order modes  have a  very similar  behaviour when  compared with
$\Delta P_{A}$ during WD cooling.}
\label{fig:he_per}
\end{figure}


\begin{figure} \caption{Kinetic energy (K.E.) vs. $T_{eff}$ for 
the same modes  with $\ell= 1$ included in  Fig. \ref{fig:he_per}. The
unit of K.E. is ergs and we have assumed that the amplitude of $y_{1}$
at surface is one. For more details, see text.}
\label{fig:he_ener}
\end{figure}


\begin{figure} \caption{Forward period spacing ($\Delta P_{k}$) vs. 
radial order, for modes with $\ell= 1$ for three models with different
$T_{eff}$ values. Symbols corresponding to modes of the same $T_{eff}$
are connected for clarity. Horizontal lines correspond to the value of
asymptotic period  spacing in each effective  temperature according to
Tassoul (1980).}
\label{fig:DeltaP}
\end{figure}


\begin{figure} \caption{The absolute value of relative difference between 
numerically  computed  and  variational  eigenperiods for  modes  with
radial  order from  $k= 1$  to $k=  56$ and  $\ell= 1$  in  all models
analyzed.}
\label{fig:variational}
\end{figure}

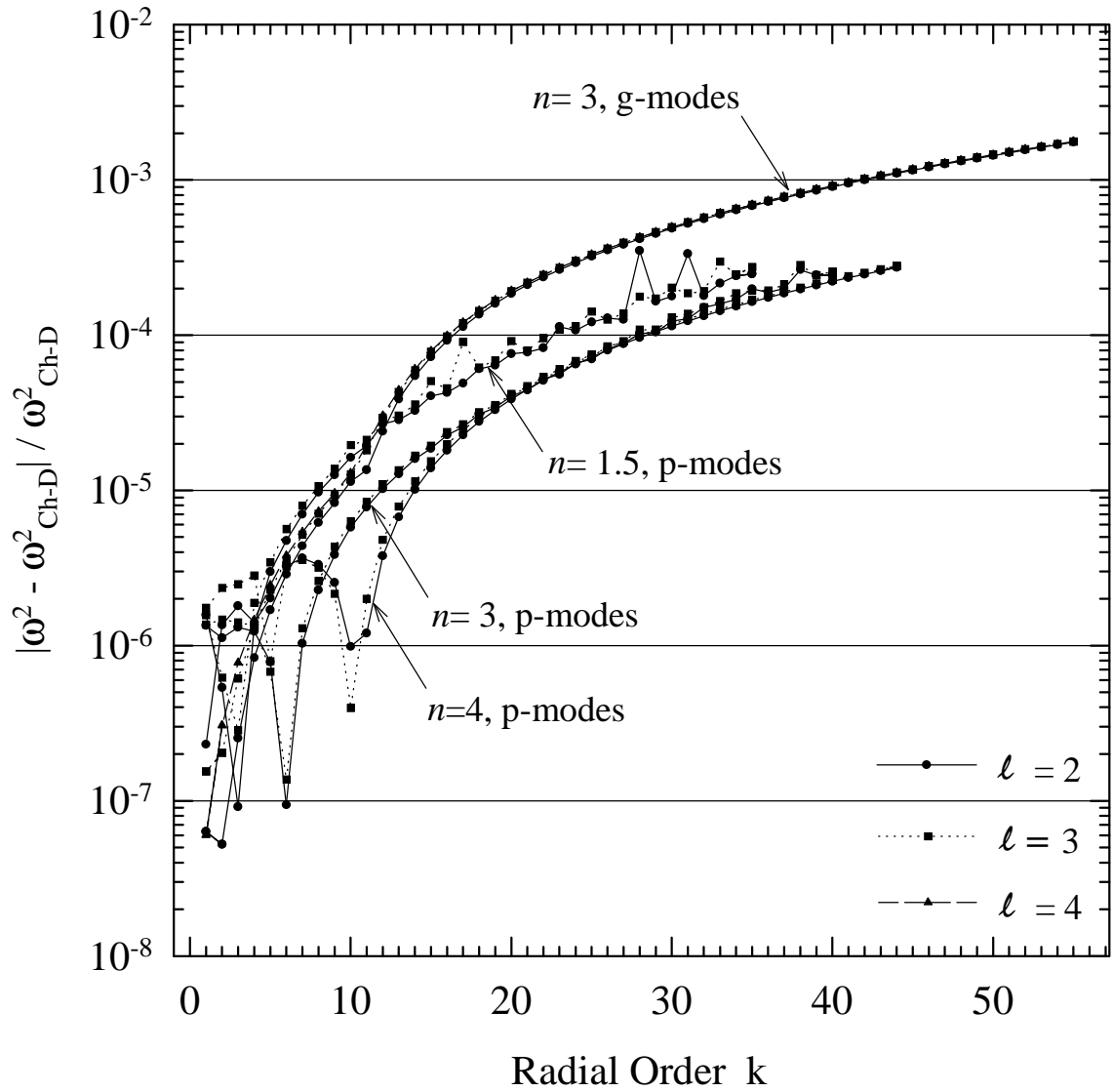
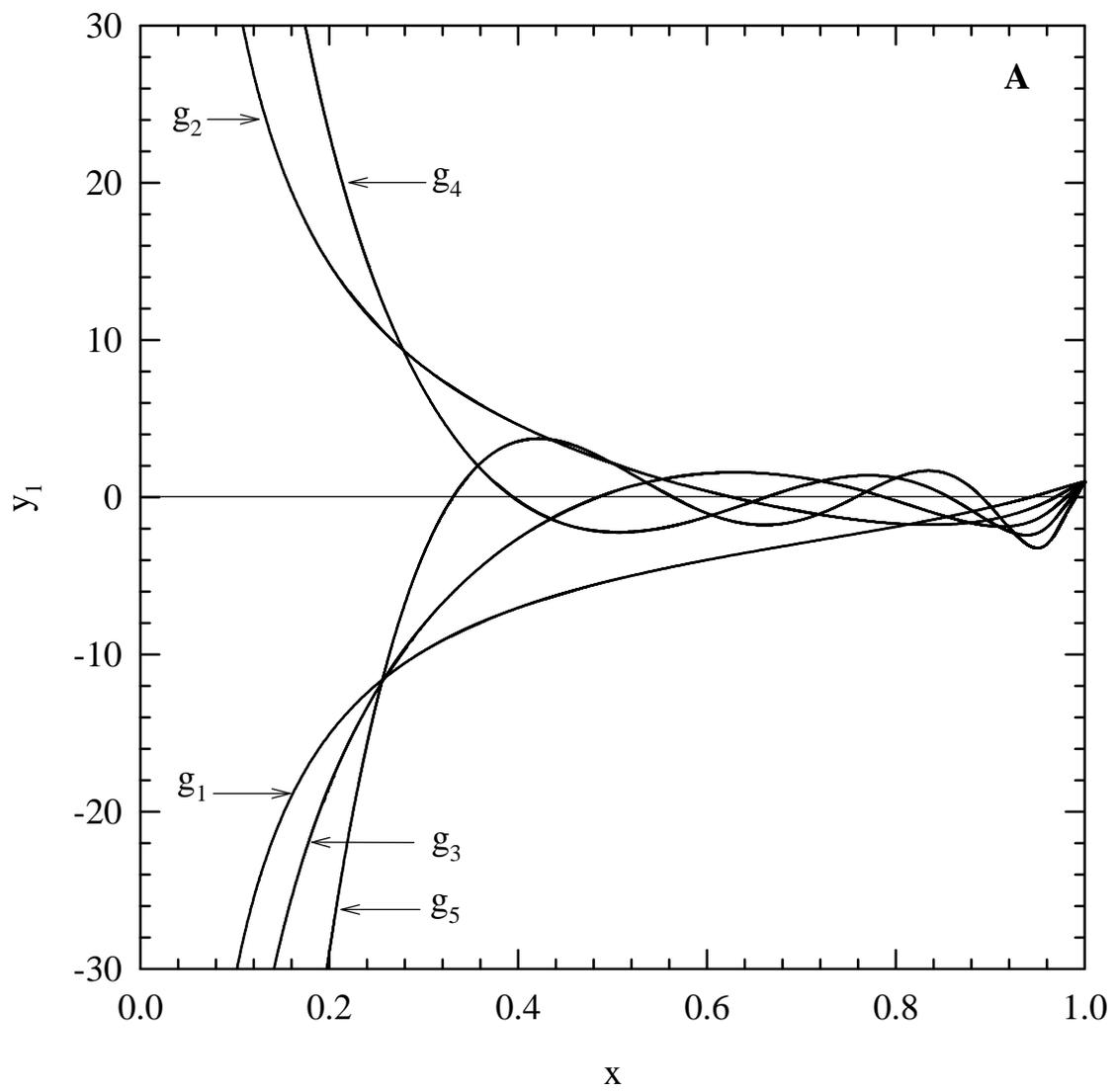
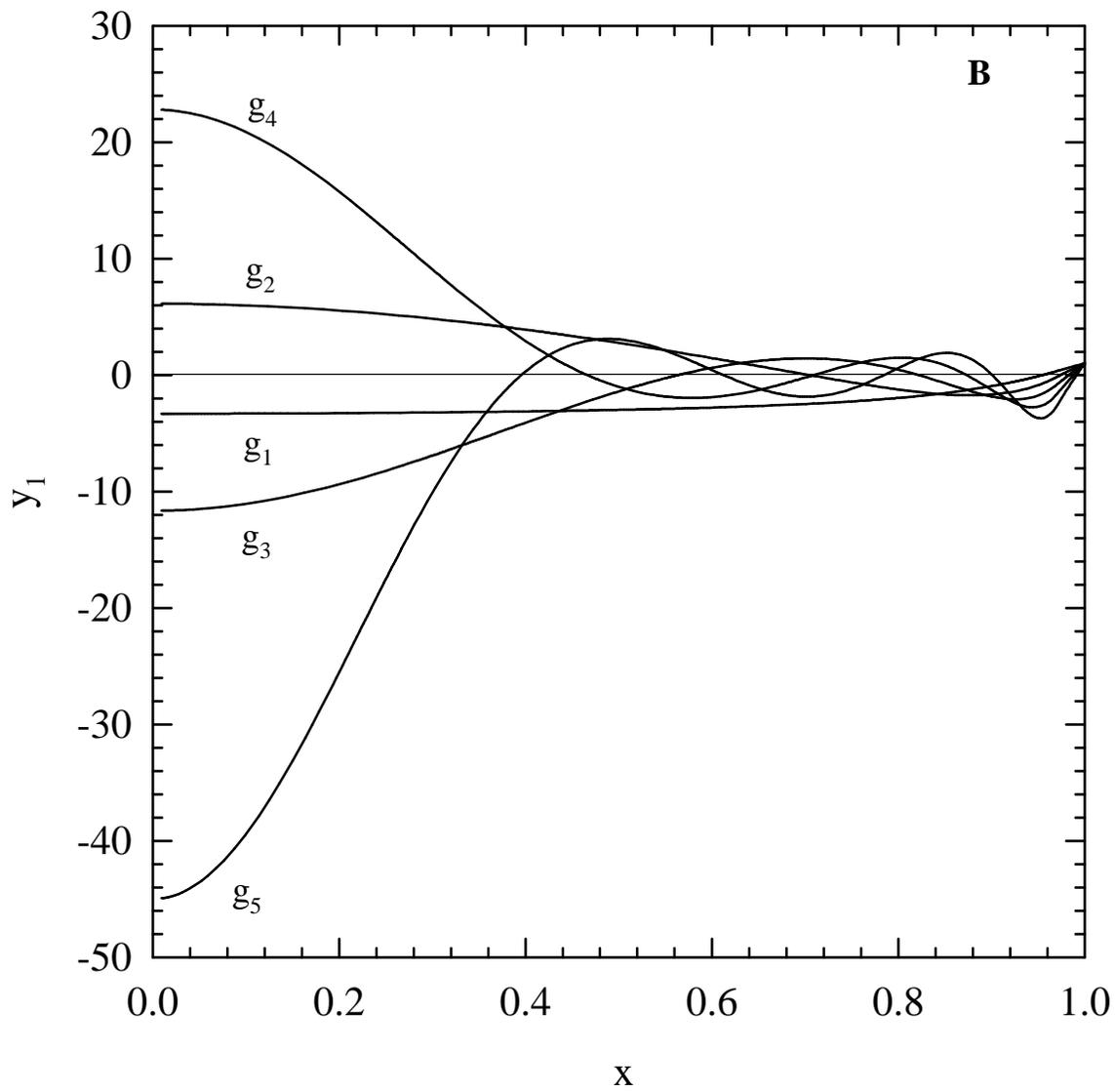
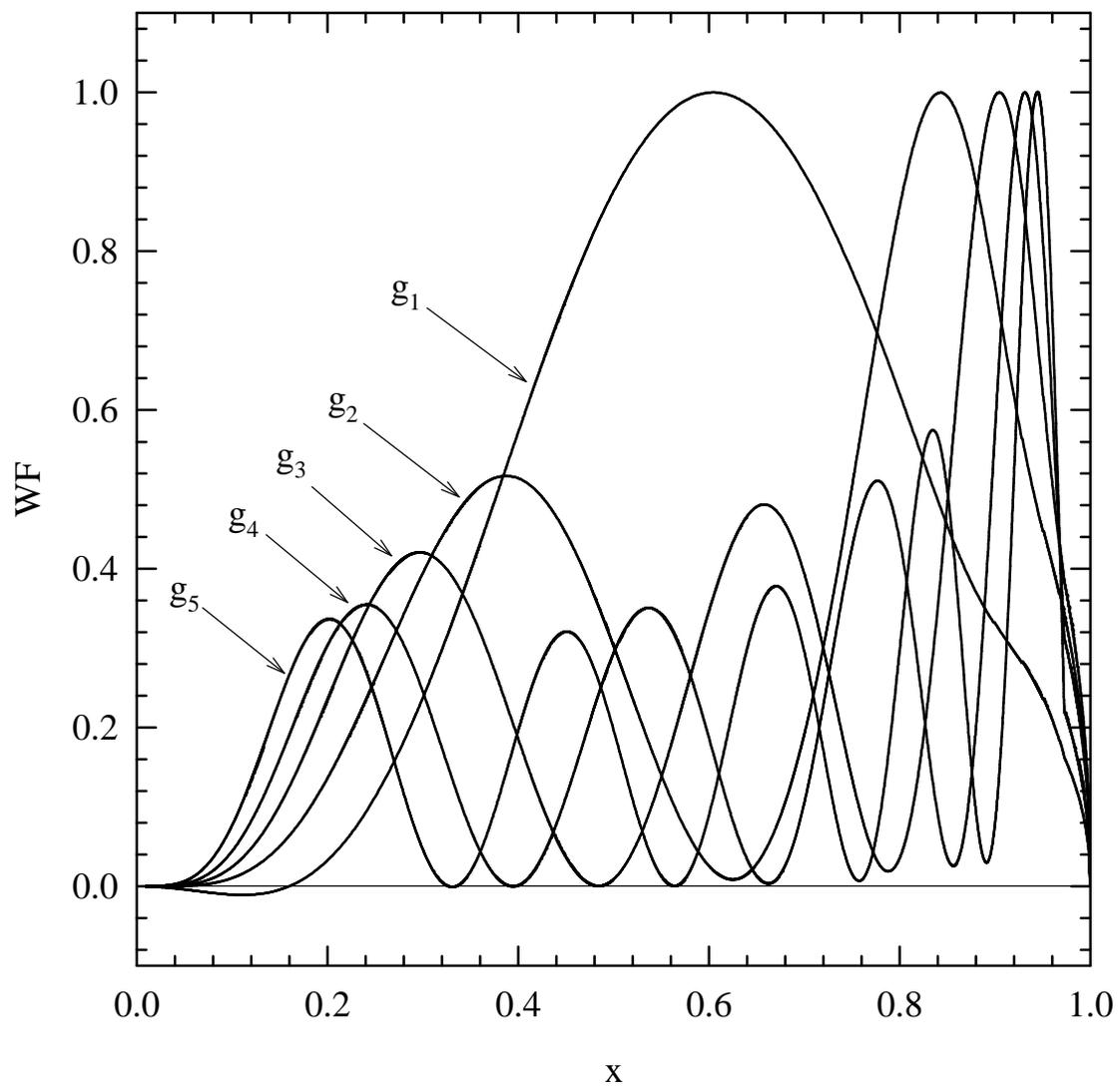
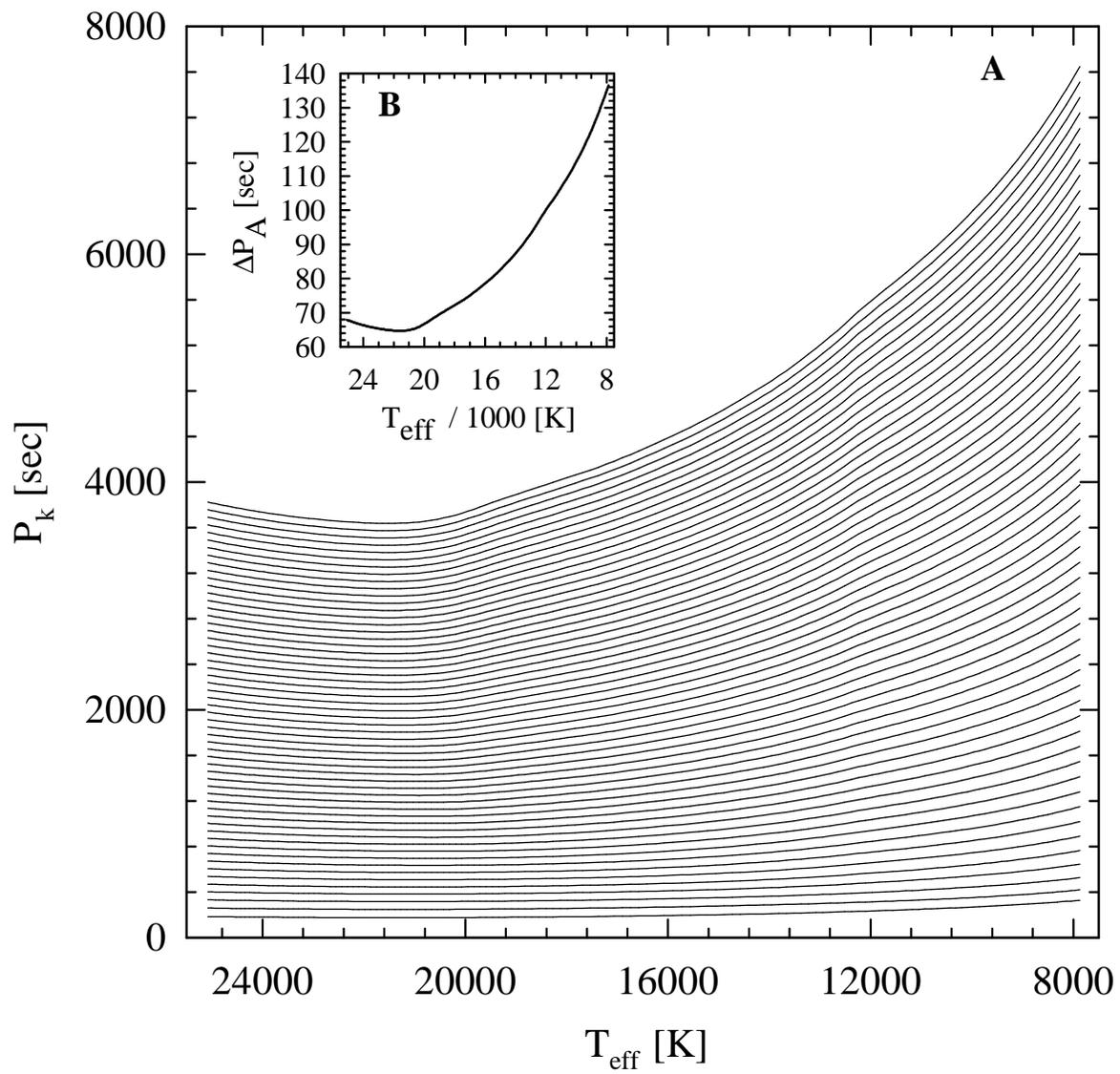
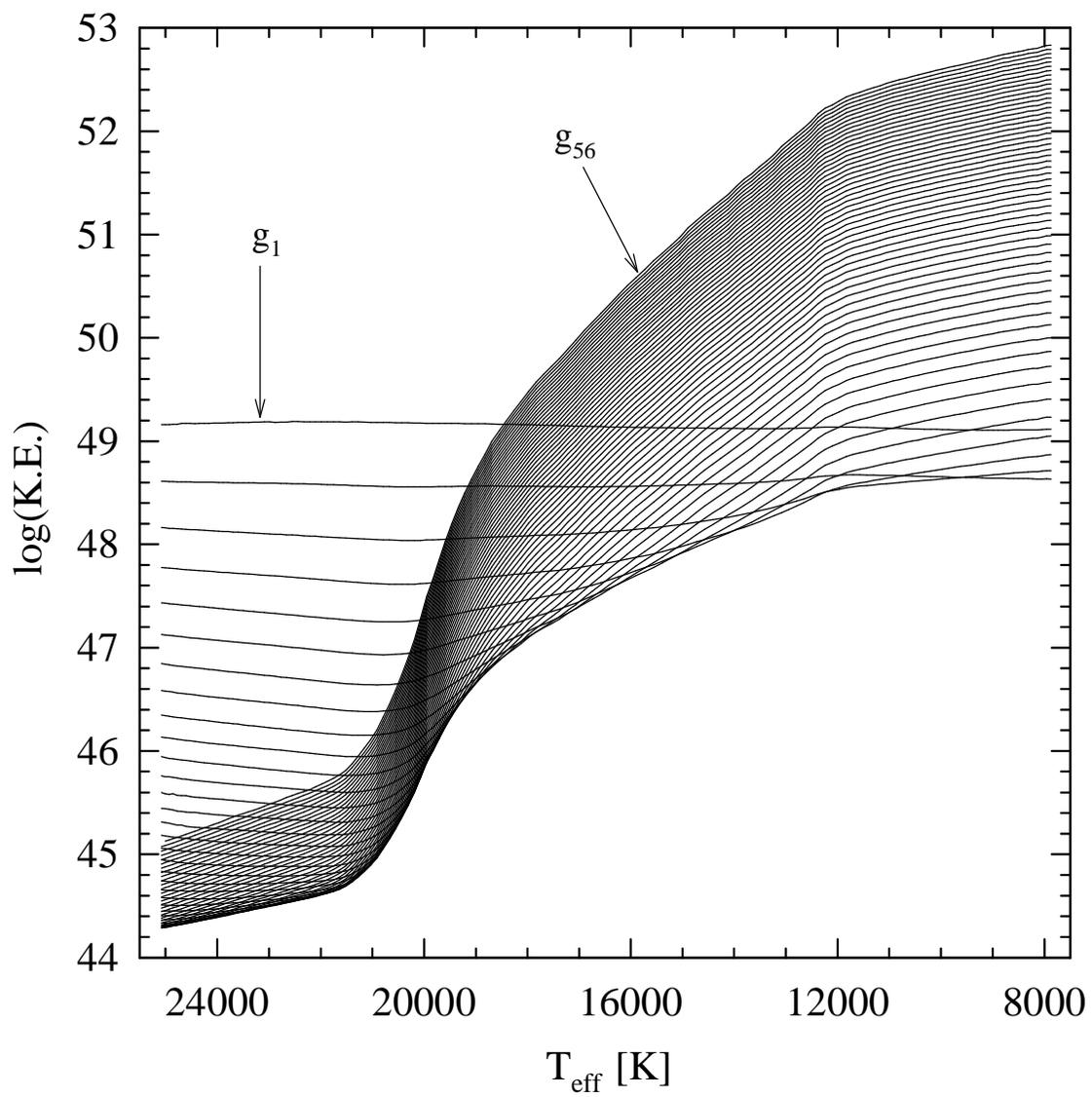
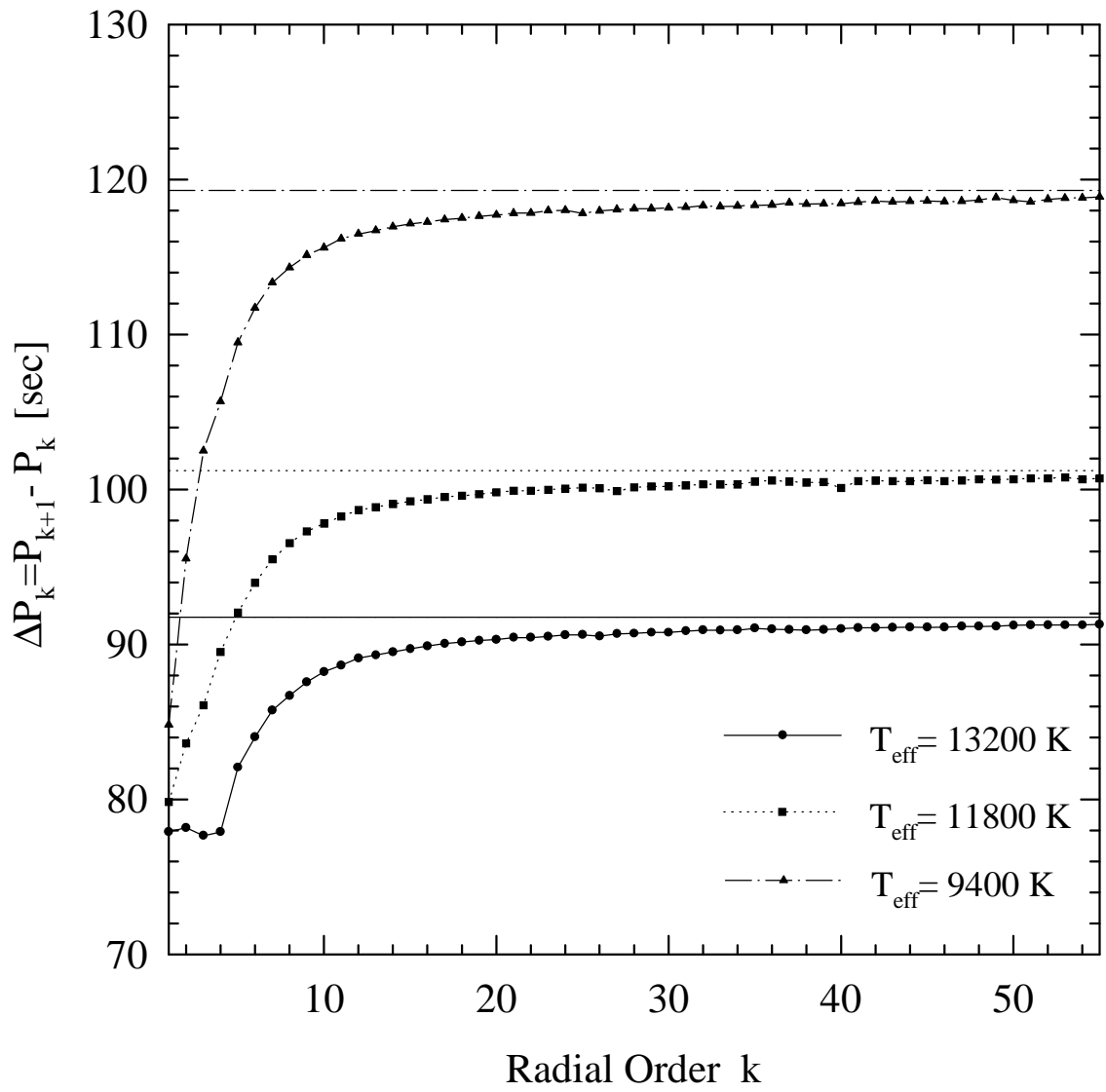
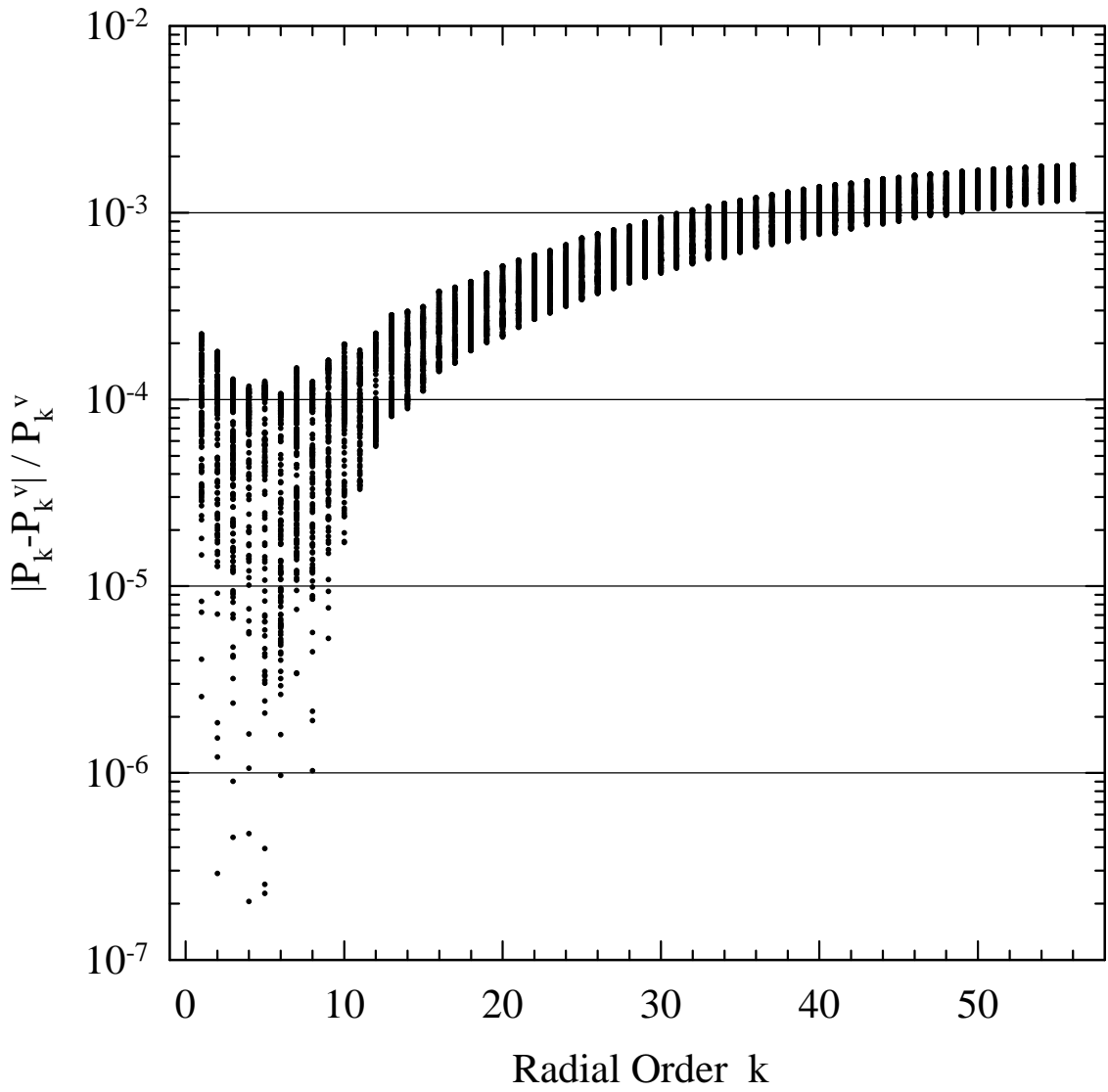
\end{document}